\documentclass[useAMS,usenatbib]{mn2e}
\usepackage{graphicx}

\def\aV{\mbox{$\rm A_V$}}

\def\rl{\mbox{$R_{\rm RDP}$}}
\def\rx{\mbox{$R_{\rm ext}$}}

\def\ms{\mbox{$M_\odot$}}
\def\ds{\mbox{$d_\odot$}}

\def\mas{\mbox{$\rm mas~yr^{-1}$}}

\def\kms{\mbox{$\rm km\,s^{-1}$}}
\def\mua{\mbox{$\mu_\alpha\cos(\delta)$}}
\def\mud{\mbox{$\mu_\delta$}}
\def\mut{\mbox{$\mu_{os}$}}
\def\sv{\mbox{$\sigma_v$}}
\def\sa{\mbox{$\sigma_\alpha$}}
\def\sd{\mbox{$\sigma_\delta$}}
\def\st{\mbox{$\sigma_{os}$}}
\def\ve{\mbox{$v_{esc}$}}

\title[Deconvolved OC-velocity dispersion]{From proper motions to star cluster 
dynamics: measuring velocity dispersion in deconvolved distribution functions}

\author[C. Bonatto and E. Bica]{C. Bonatto$^1$ and E. Bica$^1$\\
$^1$ Departamento de Astronomia, Universidade Federal do Rio Grande 
do Sul, Av. Bento Gon\c{c}alves 9500\\
Porto Alegre 91501-970, RS, Brazil}

\begin{document}

\pagerange{\pageref{firstpage}--\pageref{lastpage}}

\maketitle

\label{firstpage}

\begin{abstract}
We investigate the effect that the usually large errors associated with ground-based proper 
motion (PM) components have on the determination of a star cluster's velocity dispersion (\sv). 
Rather than histograms, we work with PM distribution functions (PMDFs), taking the $1\sigma$ 
uncertainties formally into account. In this context, a cluster's intrinsic PMDF is broadened 
by the error distribution function (eDF) that, given the average error amplitude, has a width 
usually comparable to the cluster PMDF. Thus, we apply a Richardson-Lucy (RL) deconvolution 
to the PMDFs of a set of relatively nearby and populous open clusters (OCs), using the eDFs as 
point spread functions (PSFs). The OCs are NGC\,1039 (M\,34), NGC\,2477, NGC\,2516, NGC\,2682 
(M\,67), and NGC\,7762. The deconvolved PMDFs are approximately Gaussian in shape, with 
dispersions lower than the observed ones by a factor of 4-10. NGC\,1039 and NGC\,2516, the 
nearest OCs of the sample, have deconvolved \sv\ compatible with those of bound OCs of mass 
$\sim10^3$\,\ms. NGC\,2477 and NGC\,2682 have deconvolved PMDFs with a secondary bump, shifted 
towards higher average velocities, which may be an artefact of the RL deconvolution when
applied to asymmetric profiles. Alternatively, it may originate from cluster merger, large-scale 
mass segregation or, least probably, binaries. 
\end{abstract}

\begin{keywords}
{{\em (Galaxy:)} open clusters and associations: general} 
\end{keywords}

\section{Introduction}
\label{Intro}

Star clusters continually undergo mass segregation and evaporation, tidal interactions 
with Galactic substructures, shocks with giant molecular clouds, as well as mass loss 
due to stellar evolution. By decreasing the total cluster mass - and the collective
gravitational potential, these processes affect the internal dynamics and accelerate 
the cluster dynamical evolution. Thus, the escape velocity (\ve) and, to a lesser degree, 
the space velocity dispersion (\sv), are expected to continually change with time. As a 
consequence, the majority of the open clusters (OCs) dissolve in the Galactic stellar 
field long before reaching an age of $\sim1$\,Gyr (e.g. \citealt{Lamers05}; \citealt{GoBa06}). 
In this context, the dynamical evolution of a star cluster depends critically on the balance 
between \sv\ and \ve.

For a cluster that is approximately in virial equilibrium, the space velocity dispersion 
can be computed as $\sv(\kms) = 
\sqrt{\frac{G\,M_D}{\eta\,R_{eff}}}\approx0.7\left(\frac{M_D}{10^3\ms}\right)^{1/2}
\left(\frac{R_{eff}}{1\,pc}\right)^{-1/2}$ (\citealt{Spitzer87}), where $G$ is the 
gravitational constant, $M_D$ is the dynamical mass (assumed to be stored only in single 
stars), $\eta\approx9.75$ is a constant, and $R_{eff}$ is the effective, or  projected half-light 
radius. As a scaling factor, bound clusters with $M_D\sim10^3\ms$ and $R_{eff}\sim1$\,pc 
are expected to have $\sv\sim1\,\kms$. Since the majority of the Galactic OCs have masses 
lower than $10^3$\,\ms\ (e.g. \citealt{Piskunov07}), velocity dispersions lower than, or 
of the order of $\sv\sim1\,\kms$ should be a rule. 

Obviously, the above expectation does not apply to clusters that are on their way to dissolution, 
either at the earliest evolutionary stages (less than a few $10^7$\,yr) or much later, at the remnant 
phase (several $10^8$\,yr, e.g. \citealt{PB07}). Such a super-virial state in very young clusters 
is related to the impulsive expulsion of the residual parental molecular cloud gas, driven primarily 
by the strong winds of massive stars and supernovae (\citealt{GoBa06}), thus leading to the high 
dissolution rate of young clusters (e.g. \citealt{LL2003}). Examples of dissolving young OCs with 
velocity dispersions that by far exceed the expected dynamical value (\citealt{GKG08}) are NGC\,2244, 
with an age of $\sim5$\,Myr and $\sv\sim35\,\kms$ (\citealt{CGZ07}; \citealt{N2244}), Cr\,197 and 
vdB\,92, both with an age of $\sim5$\,Myr and $\sv\sim20\,\kms$ (\citealt{vdB92})

A first-order estimate of a star cluster's \sv\ can be obtained by measuring the proper-motion 
(PM) distribution of the member stars. Having obtained the PM components in right ascension (\mua) 
and declination (\mud) of each member, the $\sa\equiv\sigma_{\mua}$ and $\sd\equiv\sigma_{\mud}$ 
dispersions can be estimated from the shape of the respective distributions. Alternatively, if
the systemic PM components are known, one can compute directly the tangential or on-sky PM 
$\mut = \sqrt{\mua^2 + \mud^2}$ and dispersion $\st=\sqrt{\sa^2 + \sd^2}$. Then, assuming isotropy, 
we can take $\sv^2 = \frac{3}{2}(\sa^2 + \sd^2)=\frac{3}{2}\st^2$.  At this point, it should
be mentioned that, contrary to radial velocity measurements, proper motions are essentially unaffected 
by binarity (\citealt{KG2008}). This occurs because radial velocities are instantaneous measurements, 
while proper motions involve a (usually large) timespan. In this sense, \citet{KG2008} show that for 
clusters of mass $\sim1000$\,\ms, binaries produce an observed radial velocity dispersion several 
times higher than the dynamical one, thus leading to an artificially-overestimated cluster mass.

Comprehensive databases such as the {\em Third U.S. Naval Observatory CCD Astrograph Catalog} 
(UCAC3, \citealt{Zach10}) and the {\em The Naval Observatory Merged Astrometric Dataset} (NOMAD,
\citealt{Zach05}) provide PM components for huge amounts of stars, either isolated or in clusters. 
However, a drawback of such all-sky, ground-based PM databases is the fact that the $1\sigma$ 
uncertainties in \mua\ and \mud\ are usually large. The effect that such errors have on the 
determination of \sv, which usually involves building PM histograms with bin sizes smaller than 
the average $1\sigma$ errors, is a constantly neglected point. Instead of histograms, our approach 
in this paper is to work with PM distribution functions (PMDFs), in which the errors are formally 
taken into account. 

This paper is organised as follows. In Sect.~\ref{PMDF} we build PMDFs and compare them
with the classical histograms. In Sect.~\ref{trgt} we select some relatively nearby and
populous OCs as test cases. In Sect.~\ref{RLD} we discuss the deconvolution method that 
we apply on the PMDFs. In Sect.~\ref{Discus} we present the results obtained so far and
discuss some observational limitations. Concluding remarks are given in Sect.~\ref{Conclu}.

\section{Proper motion analysis: histograms or distribution functions?}
\label{PMDF}

When dealing with PMs, the first step is to isolate the (most probable) cluster members,
subtract the systemic components of \mua\ and \mud\ from each star, and then compute the 
tangential or on-sky PM \mut. The systemic components can be directly computed from the 
data or obtained in OC databases (e.g. WEBDA\footnote{\em www.univie.ac.at/webda} or 
DAML02\footnote{Catalog of Optically Visible Open Clusters and Candidates 
{\em www.astro.iag.usp.br/~wilton}}). 

\begin{figure}
\resizebox{\hsize}{!}{\includegraphics{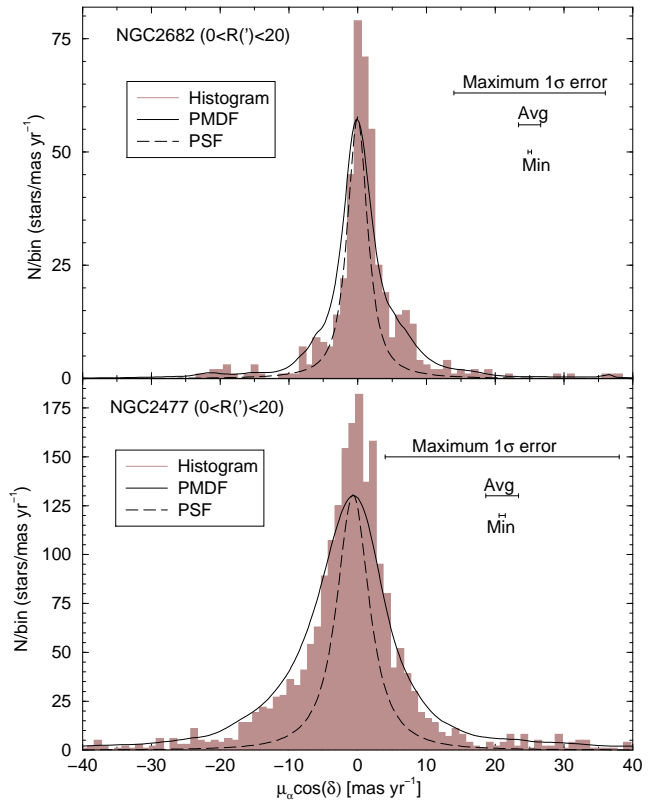}}
\caption[]{Histogram and PMDF (heavy-solid line) of NGC\,2682 (top panel) and NGC\,2477 (bottom) 
for the stars located within $R<20\arcmin$. The amplitude of the maximum, average, and minimum  
$1\sigma$ errors are illustrated (horizontal lines). Also shown is the error distribution function 
(dashed line).}
\label{fig1}
\end{figure}

Usually, the next step involves building PM histograms with a given bin width to estimate the 
dispersion either of the components or the on-sky. However, except for a few particular 
cases with high-quality astrometry\footnote{For instance, the globular cluster NGC\,6397 with 
Hubble Space Telescope WFPC2 data from 2 epochs (\citealt{Richer08}).}, ground-based PMs usually 
have significant errors, in general larger than the adopted histogram bins. Thus, by ignoring
the error amplitude, histograms should be taken only as first-order representatives of a cluster's 
intrinsic PM distribution. Wider bins would obviously minimise this effect, but would also degrade 
the PM resolution. The net result would be artificially high values of the dispersion. This is 
illustrated in Fig.~\ref{fig1} for the \mua\ component of the OCs NGC\,2682 and NGC\,2477
(Sect.~\ref{trgt}), where we show the classical histogram built with bins of size
$\Delta\mua = 1$\,\mas. Also shown are the average, minimum, and maximum values of the $1\sigma$ 
errors. Clearly, in both cases the average $1\sigma$ error corresponds to approximately the core 
($\approx4$ bins) of the histogram. It should be noted as well the introduction of a sort of
high-frequency noise, which originates from the Poisson fluctuation associated with the bin
size.

Thus, instead of histograms, the correct way of dealing with the above issue is by explicitly 
incorporating the PM component errors in PMDFs, which are defined as the fractional number of 
stars {\em per} interval of PM, $PMDF\equiv\phi(\mu)=dN/d\mu$, where $\mu$ is any component among 
\mua, \mud, and \mut. The errors are incorporated in the PMDFs by assuming that they are normally 
(i.e. Gaussian) distributed. Accordingly, if measurements of a given parameter $\chi$ are normally 
distributed around the average $\bar\chi$ with a standard deviation $\epsilon$, the probability 
of finding it at a specific value $\chi$ is given by $P(\chi)=\frac{1}{\sqrt{2\pi}\epsilon}\,
e^{{-\frac{1}{2}}\left(\frac{\chi-\bar\chi}{\epsilon}\right)^2}$. 

We start by defining a set of PM bins spanning the whole range of values observed in \mua, \mud, 
and \mut. The bins have smaller widths around the PMDF peak than for higher or lower velocities 
(for preserving profile resolution - App.~\ref{Conv}). Then, for a star with PM components and 
$1\sigma$ uncertainties $\mua\pm\epsilon_{\mua}$, $\mud\pm\epsilon_{\mud}$, $\mut\pm\epsilon_{\mut}$, 
we compute the probability of the PM of that star to be in a given bin, which is simply the difference 
of the error functions computed at the bin borders. By doing this for all stars, we end up with the
number-density of stars in each PM bin, the integral of which over the whole range of PM values is 
simply the number of stars. The \mua\ PMDFs of NGC\,2682 and NGC\,2477 are shown in Fig.~\ref{fig1}. 
By construction, the PMDFs are much smoother than the histograms (the high-frequency noise has been
naturally removed) and, because of the broadening effect due to the error spreading procedure, they 
are also somewhat wider and have a lower amplitude. Both the histograms and PMDFs are definitely 
non-Gaussian, especially because of the broad wings. 

Now, the same procedure is applied to the uncertainties, thus resulting in the intrinsic error 
distribution function (eDF). As anticipated by the amplitude of the average $1\sigma$ error, the 
eDF has a width comparable to that of the PMDF for both NGC\,2682 and NGC\,2477. In this context, 
the eDF plays a r\^ole of a PM point-spread function (PSF), which tends to broaden the intrinsic 
cluster PMDF on a degree that depends essentially on the PMDF and eDF widths. Thus, our approach
here is to deconvolve the observed PMDF of selected OCs (Sect.~\ref{trgt}) using the intrinsic eDF 
as PSF (Sect.~\ref{RLD}). 

\section{Test cases}
\label{trgt}

As test cases we searched for OCs that are located relatively nearby (for allowing detection
of low-PM components - Sect.~\ref{Discus}), away from central directions and the disk (to
minimise field-star contamination), and with a wide range of ages. Additionally, the candidates 
should have a significant number of stars with available PM components in UCAC3\footnote{Main 
features of UCAC3 are: complete sky coverage, merging of several PM catalogues, new data reduction 
with reduced errors, significantly improved photometry from CCD data, etc. See 
{\em http://cdsarc.u-strasbg.fr/viz-bin/Cat?I/315} for further details.} (for more representative 
PMDFs). The sample of OCs meeting our criteria are NGC\,1039 (M\,34), NGC\,2477, NGC\,2516, 
NGC\,2682 (M\,67), and NGC\,7762; their fundamental parameters are listed in Table~\ref{tab1}. 
In short, the selected OCs have ages within 0.2 - 4\,Gyr and distances from the Sun within 
$\sim0.4 - 1.3$\,kpc.

\begin{figure}
\resizebox{\hsize}{!}{\includegraphics{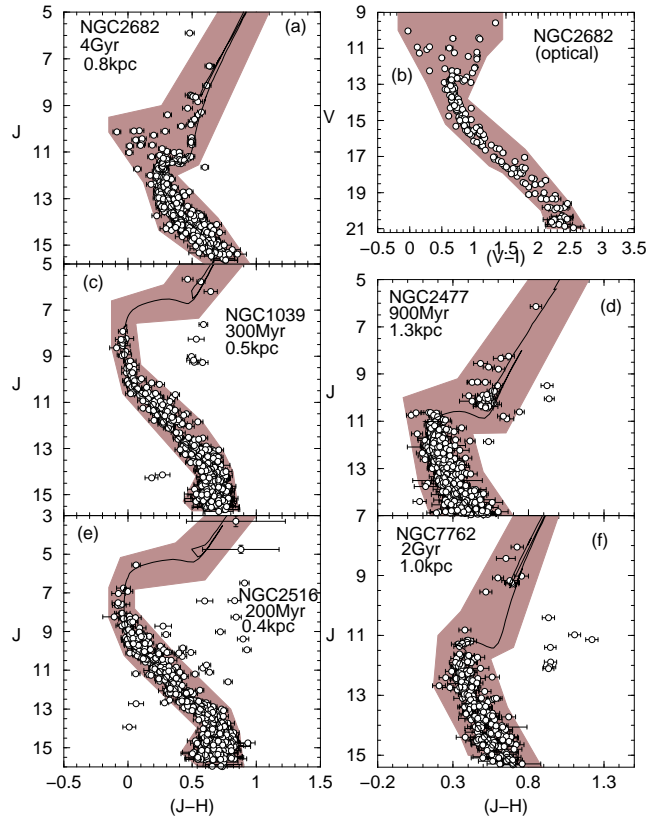}}
\caption[]{Extracted within $R<5\arcmin$, the CMDs have been built with 2MASS
photometry, except for panel (b) that was built with the VI CCD photometry of 
\citet{Yadav08}. Only stars that occur within the colour-magnitude filter (shaded
polygon) are considered in the analyses.}
\label{fig2}
\end{figure}

\begin{table*}
\caption[]{Fundamental parameters and cluster systemic PM components derived in this work}
\label{tab1}
\renewcommand{\tabcolsep}{1.1mm}
\renewcommand{\arraystretch}{1.25}
\begin{tabular}{ccccccccccccc}
\hline\hline
&\multicolumn{4}{c}{Literature}&&\multicolumn{7}{c}{This work}\\
\cline{2-5}\cline{7-13}
Cluster&$\alpha(2000)$&$\delta(2000)$&$\ell$&$b$&&Age&\aV&\ds&\rl&5\arcmin&$(\mua)_{syst}$&$(\mud)_{syst}$\\
 & (hms)&($\degr\,\arcmin\,\arcsec$)&(\degr)&(\degr)&&(Myr)&(mag)&(kpc)&(pc)&(pc)&(\mas)&(\mas)\\
(1)&(2)&(3)&(4)&(5)&&(6)&(7)&(8)&(9)&(10)&(11)&(12)\\
\hline

NGC\,1039&02:42:45&$+$42:45:42&143.66&$-$15.61&&$300\pm50$&$0.2\pm0.2$&$0.49\pm0.03$&$4.3\pm0.7$&0.7&$-0.55\pm0.01$&$-5.81\pm0.01$\\

NGC\,2477&07:52:10&$-$38:31:48&253.56&$-$5.84&&$900\pm100$&$0.8\pm0.2$&$1.31\pm0.07$&$13.3\pm1.9$&1.9&$-0.83\pm0.07$&$+1.89\pm0.06$\\

NGC\,2516&07:58:04&$-$60:45:12&273.82&$-$15.86&&$200\pm50$&$0.3\pm0.2$&$0.38\pm0.04$&$3.3\pm0.3$&0.6&$-5.60\pm0.07$&$+10.74\pm0.03$\\

NGC\,2682&08:51:18&$+$11:48:00&215.70&$+$31.90&&$4000\pm300$&$0.0\pm0.1$&$0.79\pm0.05$&$6.9\pm0.7$&1.2&$-8.61\pm0.02$&$-4.92\pm0.05$\\

NGC\,7762&23:50:01&$+$68:08:18&117.22&$+$5.85&&$2000\pm500$&$1.4\pm0.4$&$0.96\pm0.14$&$7.0\pm1.4$&1.4&$-4.67\pm0.01$&$-2.53\pm0.03$\\
         
\hline
\end{tabular}
\begin{list}{Table Notes.}
\item Cols.~4 and 5: Galactic coordinates; Col.~8: distance from the Sun; Col.~9: cluster truncation
radius (in pc); Col.~10: $R=5\arcmin$ in absolute units; Cols.~11 and 12: deconvolved systemic proper 
motion components, computed with the stars within $R<20\arcmin$.
\end{list}
\end{table*}

\begin{figure}
\resizebox{\hsize}{!}{\includegraphics{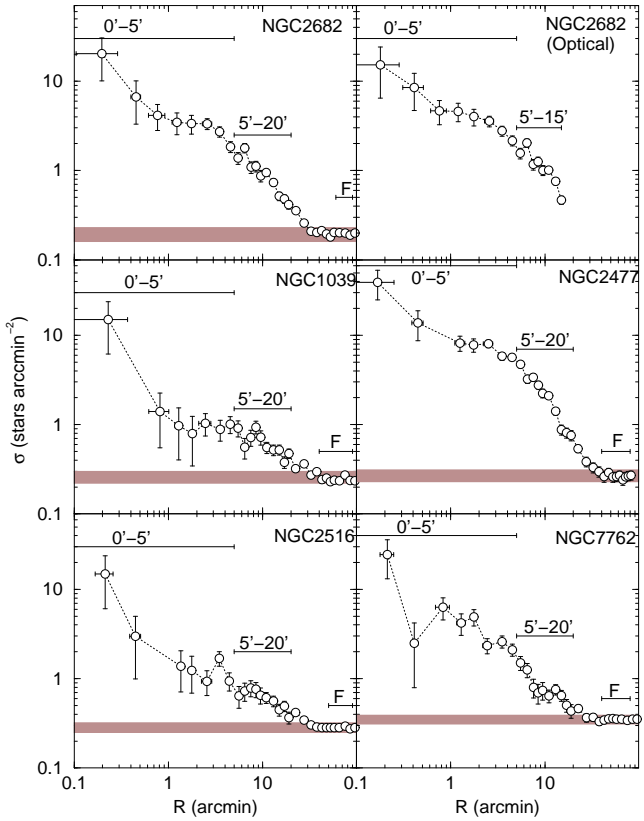}}
\caption[]{Stellar RDPs of the selected OCs. The central ($0<R(\arcmin)<5$), outer 
($5<R(\arcmin)<20$ or $5<R(\arcmin)<15$ - top right), and field (F) regions are
indicated. The field contamination level towards the central parts is illustrated
by the shaded polygon.}
\label{fig3}
\end{figure}

2MASS\footnote{The Two Micron All Sky Survey, All Sky data release - 
{\em http://www.ipac.caltech.edu/2mass/releases/allsky/}} (\citealt{2mass06}) colour-magnitude
diagrams (CMDs) extracted from the central region ($R<5\arcmin$) are shown in Fig.~\ref{fig2}. 
The 5\arcmin\ boundary was selected for practical reasons. Our analysis depends on the 
number of member stars in two different cluster regions (Sect.~\ref{RLD}), and the stellar density 
profiles (Fig.~\ref{fig3}) show that $R=5\arcmin$ represents a compromise between adequate statistics 
and a change from RDPs following a pure power-law ($R\ga5\arcmin$) to a power-law flattened by a 
core ($R\la5\arcmin$). For NGC\,2682 (M\,67) we also use 
the ground-based CCD astrometry of \citet{Yadav08} as an independent data set for cross-checking 
our results. The PM components provided by \citet{Yadav08} are already corrected for the cluster's 
systemic velocity. The 2MASS photometry was extracted from 
VizieR\footnote{\em http://vizier.u-strasbg.fr/viz-bin/VizieR?-source=II/246} in a 
wide circular field of radius $\rx=90\arcmin$, which is adequate for determining the 
cluster extension and background level (see below). We also build a colour-magnitude
filter for each cluster, which is used to isolate probable member stars (see, e.g. 
\citealt{Teu34} for a discussion on the use of such filters). Only stars with colour 
and magnitude compatible with the filter are used in the subsequent analyses. Age,
distance from the Sun, and reddening are derived by fitting solar-metallicity Padova 
isochrones (\citealt{Girardi2002}; \citealt{TheoretIsoc}) to the CMDs. The derived
fundamental parameters (Table~\ref{tab1}) are similar to those in WEBDA.

\begin{figure}
\resizebox{\hsize}{!}{\includegraphics{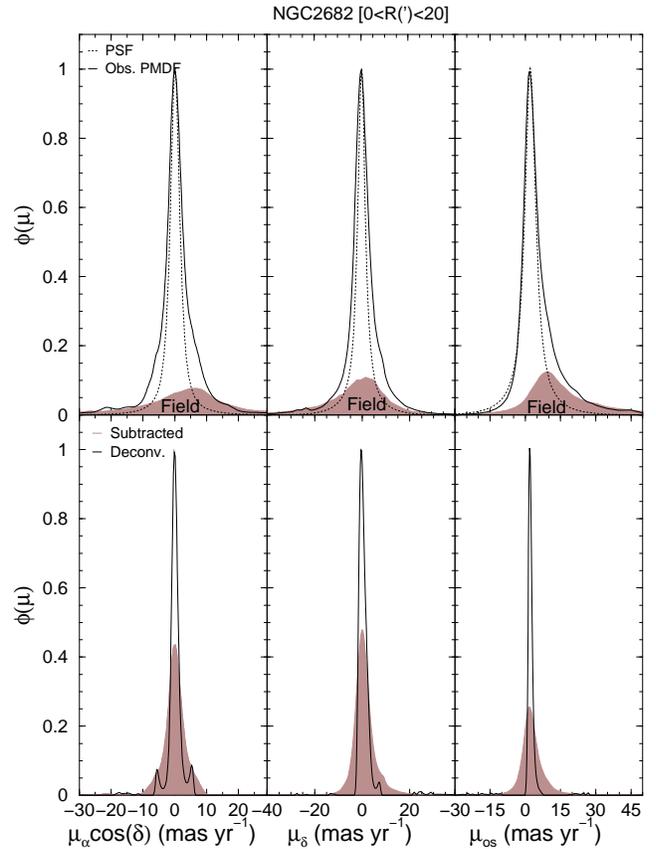}}
\caption[]{RL deconvolution applied to the \mua\ (left panel), \mud\ (centre), and
\mut\ (right) PM components for the stars located in the region $R<20\arcmin$ of 
NGC\,2682. Top panels show the observed cluster (solid line) and field (shaded area) 
PMDFs, and the PSF (dotted line). Bottom: field-subtracted (shaded area) and deconvolved 
(solid line) PMDFs.}
\label{fig4}
\end{figure}

After isolating the probable member stars with the colour-magnitude filters, we use them 
to build the stellar radial density profile (RDP), which is the projected stellar number 
density profile around the cluster centre (Fig.~\ref{fig3}).  Note that the cluster
centres have been computed by an algorithm that searches for the highest stellar density 
in the innermost bin and, at the same time, the smoothest RDP (\citealt{vdB92}). Working
with colour-magnitude filtered photometry minimises field contamination and enhances the
RDP contrast with the fore/background. The position (and error) of each RDP point along 
the $R$ axes in Fig.~\ref{fig3} corresponds to the average (and $1\sigma$ uncertainty) 
distance to the cluster centre of the stars within each bin. We also estimate the cluster
truncation radius (\rl), which is simply the distance from the cluster centre where RDP
and background are statistically indistinguishable (Fig.~\ref{fig3}). On average, the 
boundary $R=5\arcmin$ represents $\approx1/6$ of \rl\ (Table~\ref{tab1}). The near-infrared RDPs
show that the selected OCs have an extension reaching $\approx20-30\arcmin$ (the optical
data of \citealt{Yadav08} are restricted to $\approx15\arcmin$). They also show that
field stars with the same colour and magnitude as the probable members are still present
in the central region, in varying amounts for the different OCs. The residual field 
contamination will be taken into account in the PM analysis (Sect.~\ref{RLD}).

Proper motion components were obtained in UCAC3 (that also provides the 2MASS photometry
for each star) based on the same central coordinates and extraction radius as those used 
for building the CMDs and RDPs. Afterwards, we applied the respective colour-magnitude 
filters (Fig.~\ref{fig2}) before computing the PM distributions (see below). 

\section{PMDF deconvolution}
\label{RLD}

By construction, the observed PMDFs are broadened by a PSF that corresponds to the eDF 
(Sect.~\ref{PMDF}). Thus, to uncover the intrinsic PMDFs, we apply the iterative 
Richardson-Lucy (RL) deconvolution method proposed by \citet{Richard72} and \citet{Lucy74}.
The RL deconvolution conserves the PMDF integral (in the present context the number of 
stars), but has a relatively slow convergence rate.

We illustrate the RL deconvolution with the \mua, \mud, and \mut\ components of the OC 
NGC\,2682 (Fig.~\ref{fig4}). As the first step we define the PM-bin size distribution
(App.~\ref{Conv}), and compute and subtract the systemic PM components (Table~\ref{tab1}) 
from \mua\ and \mud. Actually, this is an iterative step, in which we first subtract the 
field contamination, deconvolve the resulting PMDF and obtain the systemic PM. We use the 
region $R<20\arcmin$ in the analysis as a compromise between the number of member stars 
and contrast with the field level (Fig.~\ref{fig3}). Then we build the cluster and field 
PMDFs, together with the eDF, which has a width similar to that of the cluster PMDF. Next, 
the contaminant PMDF of the residual field stars, which is especially seen in \mua\ as a 
bump centred at $\approx6\,\mas$, is subtracted from the observed PMDFs. Finally, the RL 
deconvolution is applied to the field-subtracted PMDF.

\begin{figure}
\resizebox{\hsize}{!}{\includegraphics{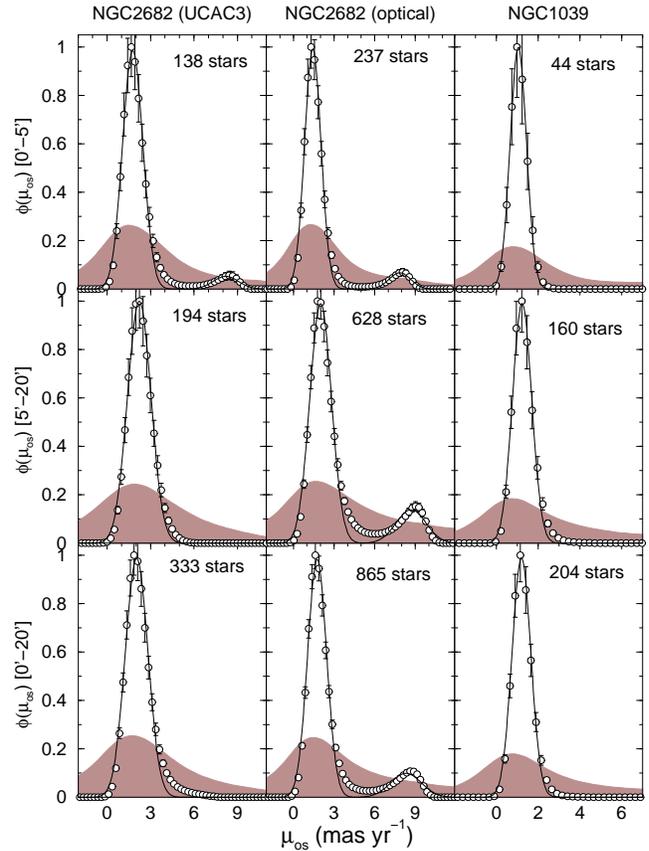}}
\caption[]{Field-subtracted (shaded area) and deconvolved (circles) on-sky PMDFs 
built for different cluster regions: central (top), outer (middle) and overall (bottom). 
The solid line shows the Gaussian fit to the deconvolved PMDF. The number of member stars 
in each region is indicated. The PMDFs with the optical data of \citet{Yadav08} for NGC\,2682 
(central panels) are restricted to $R<15\arcmin$.}
\label{fig5}
\end{figure}

\begin{figure}
\resizebox{\hsize}{!}{\includegraphics{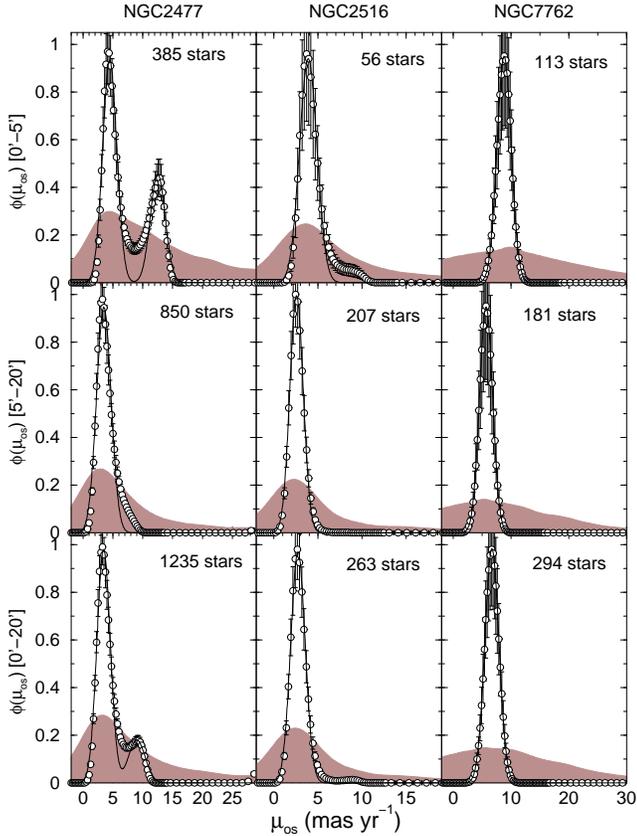}}
\caption[]{Same as Fig.~\ref{fig5} for the remaining clusters.}
\label{fig6}
\end{figure}

Besides the slow convergence rate, the RL deconvolution is also known for not having 
a universal convergence criterion. Thus, for determining the number of deconvolution
iterations ($N_{it}$), we simply sum the squared difference between successive iterations 
over all bins of the deconvolved PMDF, defined as $$\chi_j^2=\sum_i{\left\{PMDF(x_i)_j -
PMDF(x_i)_{j-1}\right\}^2},$$ where $j$ is the current iteration and $x_i$ is the 
$i^{th}$ bin along the PM axis (either \mua, \mud, or \mut), and compute the fractional 
variation of $\chi$, $f_{\chi} = 1-\chi_j/\chi_{j-1}$. After a series of tests to check 
changes in deconvolution parameters with the number of iterations, we arbitrarily decided 
to stop when$N_{it}=200$ (or $f_{\chi}=0.003-0.005$; App.~\ref{Conv}). Under this criterion, 
the deconvolved PMDFs end up with significantly lower dispersions than the observed ones 
(Figs.~\ref{fig4} to \ref{fig6} and Table~\ref{tab2}), without significant added noise 
or artifacts (App.~\ref{Conv}). Besides, contrary to the observed PMDFs, the shape of 
the deconvolved ones is approximately Gaussian. 

\begin{table*}
\caption[]{On-sky proper motion components}
\label{tab2}
\renewcommand{\tabcolsep}{2.4mm}
\renewcommand{\arraystretch}{1.0}
\begin{tabular}{ccccccccccccc}
\hline\hline
&&&\multicolumn{2}{c}{Observed}&&\multicolumn{5}{c}{Deconvolved}\\
\cline{4-5}\cline{7-11}
Region&N&&$\bar\mu$&$\sigma$&&$\bar\mu$&$\sigma$&&$\bar\mu$&$\sigma$&&$\sigma_{obs}/\sigma_{dec}$\\
(\arcmin)&(stars)&&(\mas)&(\mas)&&(\mas)&(\mas)&&(\kms)&(\kms)\\
(1)&(2)&&(3)&(4)&&(5)&(6)&&(7)&(8)&&(9)\\
\hline
\multicolumn{13}{c}{NGC\,1039 --- $\rm age\sim300$\,Myr, $\ds\sim0.5$\,kpc}\\
\hline
0---20&204&& $\approx1.2$ & $\ga2.4$ &&$1.20\pm0.04$&$0.45\pm0.04$&&$2.8\pm0.1$&$1.0\pm0.1$&&$\ga5.1$\\
5---20&160&& $\approx1.2$ & $\ga2.5$ &&$1.23\pm0.04$&$0.46\pm0.04$&&$2.9\pm0.1$&$1.1\pm0.1$&&$\ga5.2$\\
0---5 &44&&  $\approx1.0$ & $\ga1.9$ &&$1.05\pm0.04$&$0.39\pm0.04$&&$2.4\pm0.1$&$0.9\pm0.1$&&$\ga4.9$\\
\hline
\multicolumn{13}{c}{NGC\,2477 --- $\rm age\sim900$\,Myr, $\ds\sim1.3$\,kpc}\\
\hline
0---20&1030&& $\approx3.3$ & $\ga5.3$ &&$3.44\pm0.05$&$1.12\pm0.05$&&$21.4\pm0.3$&$6.9\pm0.3$&&$\ga4.6$\\
0---20&205 && ---          & ---      &&$9.00\pm0.50$&$1.40\pm0.40$&&$55.9\pm3.1$&$8.7\pm2.5$&&---\\
5---20&850 && $\approx2.8$ & $\ga4.8$ &&$3.48\pm0.04$&$1.22\pm0.04$&&$21.7\pm0.3$&$7.6\pm0.3$&&$\ga3.9$\\
0---5 &232 && $\approx4.5$ & $\ga7.5$ &&$4.47\pm0.05$&$1.19\pm0.05$&&$27.8\pm0.3$&$7.4\pm0.3$&&$\ga6.0$\\
0---5 &153 && ---          & ---      &&$12.5\pm0.5$ &$1.25\pm0.25$&&$77.6\pm3.1$&$7.8\pm1.6$&&---\\
\hline
\multicolumn{13}{c}{NGC\,2516 --- $\rm age\sim200$\,Myr, $\ds\sim0.4$\,kpc}\\
\hline
0---20&263&& $\approx2.7$ & $\ga3.6$ &&$2.72\pm0.02$&$0.82\pm0.02$&&$4.9\pm0.1$&$1.5\pm0.1$&&$\ga4.3$\\
5---20&207&& $\approx2.5$ & $\ga3.4$ &&$2.56\pm0.02$&$0.76\pm0.02$&&$4.6\pm0.1$&$1.4\pm0.1$&&$\ga4.5$\\
0---5 &56 && $\approx4.1$ & $\ga4.2$ &&$3.89\pm0.02$&$0.99\pm0.02$&&$7.0\pm0.1$&$1.8\pm0.1$&&$\ga4.2$\\
\hline
\multicolumn{13}{c}{NGC\,2682 (UCAC3) --- $\rm age\sim4$\,Gyr, $\ds\sim0.8$\,kpc}\\
\hline
0---20&333&& $\approx2.1$ & $\ga3.2$ &&$2.04\pm0.02$&$0.78\pm0.03$&&$7.6\pm0.1$&$2.9\pm0.1$&&$\ga4.1$\\
5---20&194&& $\approx2.2$ & $\ga3.5$ &&$2.22\pm0.03$&$0.81\pm0.03$&&$8.3\pm0.1$&$3.0\pm0.1$&&$\ga4.3$\\
0---5 &129&& $\approx1.9$ & $\ga2.8$ &&$1.77\pm0.03$&$0.70\pm0.03$&&$6.6\pm0.1$&$2.6\pm0.1$&&$\ga4.0$\\
0---5 &9  && ---          & ---      &&$8.20\pm0.24$&$0.88\pm0.24$&&$30.7\pm0.9$&$3.3\pm0.9$&&---\\
\hline
\multicolumn{13}{c}{NGC\,2682 (optical)}\\
\hline
0---15&752&& $\approx1.4$ & $\ga3.4$ &&$1.75\pm0.02$&$0.70\pm0.02$&&$6.6\pm0.1$&$2.6\pm0.1$&&$\ga4.9$\\
0---15&113&& ---          & ---      &&$8.32\pm0.17$&$1.13\pm0.17$&&$31.1\pm0.6$&$4.2\pm0.6$&&---\\
5---15&535&& $\approx1.7$ & $\ga4.1$ &&$2.01\pm0.03$&$0.80\pm0.03$&&$7.5\pm0.1$&$3.0\pm0.1$&&$\ga5.1$\\
5---15&93 && ---          & ---      &&$8.69\pm0.15$&$1.10\pm0.15$&&$32.5\pm0.6$&$4.1\pm0.6$&&---\\
0---5 &217&& $\approx1.3$ & $\ga2.4$ &&$1.43\pm0.02$&$0.61\pm0.02$&&$5.4\pm0.1$&$2.3\pm0.1$&&$\ga4.0$\\
0---5 &20 && ---          & ---      &&$7.86\pm0.16$&$0.80\pm0.16$&&$29.4\pm0.6$&$3.0\pm0.6$&&---\\
\hline
\multicolumn{13}{c}{NGC\,7762 --- $\rm age\sim2$\,Gyr, $\ds\sim1.0$\,kpc}\\
\hline
0---20&294&& $\approx6.4$ & $\ga11.8$ &&$6.70\pm0.02$&$1.22\pm0.02$&&$30.1\pm0.1$&$5.5\pm0.1$&&$\ga9.7$\\
5---20&181&& $\approx5.5$ & $\ga11.2$ &&$5.69\pm0.02$&$1.13\pm0.02$&&$25.6\pm0.1$&$5.1\pm0.1$&&$\ga10$\\
0---5 &113&& $\approx8.5$ & $\ga12.7$ &&$8.81\pm0.02$&$1.23\pm0.02$&&$39.6\pm0.1$&$5.5\pm0.1$&&$\ga10$\\
\hline
\end{tabular}
\begin{list}{Table Notes.}
\item Col.~2: number of member stars in region. $\bar\mu$ and $\sigma$ derived from
the fit $PMDF(\mu)\propto e^{-0.5\left((\mu-\bar\mu)/\sigma\right)^2}$. Conversion 
from $\rm mas\,yr^{-1}$ to \kms\ was based on the respective cluster distances 
(Table~\ref{tab1}).
\end{list}
\end{table*}

\section{Discussion} 
\label{Discus}

Similarly to NGC\,2682 (Sect.~\ref{RLD}), the RL deconvolution was applied to the field-subtracted 
PMDFs of the remaining OCs. Here we present the results of this procedure but, to avoid redundancy, 
we restrict the discussion to the on-sky PMDFs. However, besides the overall cluster, we also analyse 
separately the central and outer regions (according to the RDPs in Fig.~\ref{fig3}), to search for 
spatial variations in the stellar kinematics that may be related to dynamical evolution. The deconvolved 
PMDFs are subsequently fitted with a Gaussian profile 
($PMDF(\mu)\propto e^{-0.5\left((\mu-\bar\mu)/\sigma\right)^2}$), which provides the velocity dispersion 
($\sigma$) and the average velocity ($\bar\mu$) of the stars in the region. The field-subtracted and 
deconvolved PMDFs are shown in Figs.~\ref{fig5} and \ref{fig6}, and the relevant profile parameters 
are given in Table~\ref{tab2}. The observed, field, and error PMDFs for the full OC sample are shown 
in App.~\ref{OFP}.

It is clear that the deconvolved PMDFs (with dispersion $\sigma_{dec}$) are significantly narrower 
than the field-subtracted ones ($\sigma_{obs}$). Indeed, we find $\sigma_{dec}\approx(0.1-0.25)\,\sigma_{obs}$, 
but this ratio should be taken as an upper limit, since the observed PMDFs are not Gaussian and the fit 
is dominated by the profile core.

The central and outer parts of NGC\,1039, NGC\,2516, and NGC\,7762, have PMDFs
essentially characterised by a single Gaussian. In some cases (NGC\,2682, NGC\,2477, and only 
marginally in NGC\,2516) however, the observed PMDF has a wing towards high-PM values, which 
appears to be a signature of structure in the profile. This structure shows up in the deconvolved 
PMDFs as a second PM component, which is conspicuous especially in the central cluster region. 
The second component is characterised by a higher average velocity than the main one, 
but both have similar values of dispersion. 

In the central parts of NGC\,2477, the second component corresponds to about 40\% of the 
member stars in the region. This fraction drops to $\sim17\%$ when the overall cluster is 
considered. For NGC\,2682, this component contains only $\sim7\%$ of the stars in the central
region. This fraction is consistently the same for the UCAC3 and optical data. However, in the
outer parts of NGC\,2682, the second component is only seen in the optical data. Possible reasons 
for the difference are: the optical PMDF corresponds to a smaller and somewhat more interior region 
(5\arcmin - 15\arcmin\ as compared to 5\arcmin - 20\arcmin), and contains $\sim3$ times as much 
stars as that of UCAC3. The bump appears only when the observed PMDF is clearly non-asymmetric 
with respect to the average velocity, thus displaying a broad wing towards high velocities, e.g. 
NGC\,2477 and NGC\,2682. On the other hand, there is no bump emerging from the essentially gaussian 
(observed) PMDFs of NGC\,1039, NGC\,2516, and NGC\,7762 (Figs.~\ref{fig5} and \ref{fig6}). In
addition, it is interesting to note that the bump in the central region of NGC\,2682 appears 
almost identically in PMDFs built with independent data sets. 

Another issue is to what degree incompleteness in crowded regions - and the more difficult measurement 
of PM components for faint stars - affect the PMDFs. Additionally, could the bump be related to 
incompleteness? We use NGC\,2477, the most distant and populous OC of our sample (thus, the most 
prone to suffering from incompleteness) to examine this point (App.~\ref{dMagBump}). Since our analysis
depends on the number of member stars (especially at the central region), we selected $J=12.5$ as the 
boundary between bright and faint stars. At the distance of NGC\,2477, this boundary corresponds to a 
stellar mass of $m\approx1.6\,\ms$. The bright and faint deconvolved PMDFs built for the central region 
are similar (Fig.~\ref{fig11}), consistently presenting the high-velocity bump at $\bar\mu\approx12\,\mas$. 
The only significant difference is that the faint (less massive) star PMDF is shifted $\sim1.3\mas$ towards 
high values of $\bar\mu$ with respect to that of the bright stars. A similar shift occurs for the bright 
and faint star PMDFs in the outer region (without the bump). Although relatively small, the shift 
$\Delta\bar\mu\approx1.3\,\mas\approx8\,\kms$ between the bright and faint PMDFs might suggest slightly 
different kinematics for stars in different mass ranges. This experiment also suggests that incompleteness 
and PM measurements of faint stars - at least to the level available in UCAC3 - are not critical for the 
PMDFs. 

Additionally, one might ask whether the bump may come from residual, i.e. unaccounted for 
field contamination. It is true that, given the statistical way we {\em decontaminate} the clusters 
(Sect.~\ref{trgt}), some field-star contribution might persist in the subtracted PMDFs. However, both 
NGC\,2477 and NGC\,2782 are located in the third Galactic quadrant and at high Galactic latitudes, which 
by itself minimises contamination (as can also be seen in Figs.~\ref{fig9} and \ref{fig10}). Thus, any 
residual contamination should be minimum, which would contradict the fraction of stars composing the bump, 
$\approx40\%$ in the central region of NGC\,2477 and $\approx7\%$ in NGC\,2682. The above arguments suggest 
that the bump is related to a cluster kinematic property, but we cannot definitely rule out the possibility 
that it might be an artefact of the RL deconvolution, and/or some residual contamination by stars with 
peculiar PM components.

There is no direct interpretation for the additional bump seen in the deconvolved PMDFs of 
NGC\,2477 and NGC\,2682. Assuming that it is physical, one possibility is that the double peak 
may arise from a merger of two clusters (as suggested by the referee,  Thijs Kouwenhoven). In 
this sense, \citet{OBD2000} carried out N-body simulations of cluster encounters, studying long-term 
structural changes up to $\sim1$\,Gyr. They found that the clusters may coalesce at such ages, 
but until then, the presence of the two clusters can still be traced by means isophotal distortions 
and ellipticity variations, as observed in model and actual clusters (e.g. \citealt{ODBD2000}). In
this scenario, internal differences in kinematics might persist too, producing different signatures
in the deconvolved PMDFs.

Alternatively, the high-average velocity component may be related to mass segregation, in which 
a fraction of the stars collectively migrate along the radial direction (over a time-scale of a 
relaxation time) within a star cluster. In the cases dealt with here, it occurs only in the two 
most populous OCs, NGC\,2477 and NGC\,2682. Maybe it cannot be detected (by the present approach) 
in the other OCs because they are less populated (i.e., possibly the same reason why it is seen in 
the outer parts of NGC\,2682 with the optical data, but not with UCAC3). Finally, working with 
histograms and a different PM-data set, \citet{M67} raised the possibility that the high-velocity 
component in NGC\,2682 might be related to the presence of binaries. However, given the findings
of \citet{KG2008}, this possibility seems the least probable. In any case, a definitive solution
for the bump nature would require detailed simulations of the internal cluster dynamics (including
cluster merger, mass segregation, and binarity), a task that is beyond the scope of the present 
paper.


\begin{figure}
\resizebox{\hsize}{!}{\includegraphics{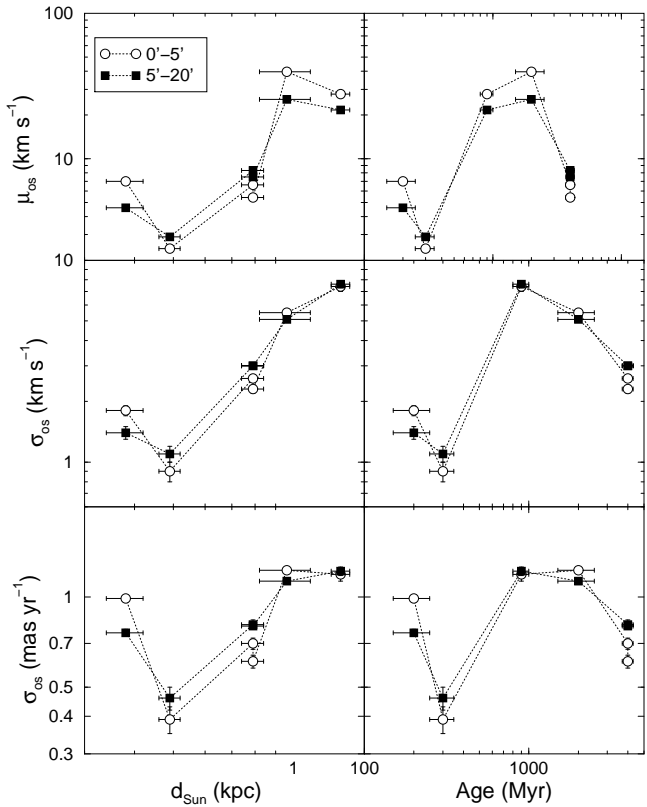}}
\caption[]{Dependence of the deconvolved profile parameters on distance from the Sun (left panels) 
and cluster age (right). The central (empty circles) and outer (filled squares) regions are shown. 
Given the rather limited range of values of $\sigma_{obs}/\sigma_{dec}$ (Table~\ref{tab2}), similar 
relations hold as well for the observed values of $\bar\mu$ and \st.}
\label{fig7}
\end{figure}

The velocity dispersions derived from the deconvolved PMDFs (Table~\ref{tab2}) of NGC\,1039 
and NGC\,2516 (and, to a lesser degree, NGC\,2682), are consistent with those of (approximately
virialized) OCs of $\sim10^3$\,\ms. On the other hand, those of NGC\,2516 and NGC\,2477 appear 
to be excessively high for OCs of a similar mass scale. However, such large dispersion values 
can be partly explained by an observational limitation related to distance, since both OCs are 
the most distant of the sample. For a limited observational time-base, the PM determination for 
a distant OC will primarily detect the high-PM components, thus leading to a broad PM profile 
and high values of velocity dispersion. Under similar conditions, a nearby OC, on the other 
hand, will also have part of the low-PM components detected, thus implying a lower velocity 
dispersion. This effect is present in our analysis, as can be seen by the correlation between 
\st\ and distance from the Sun (Fig.~\ref{fig7} - left panels). An additional consequence of 
the distance-related effect would be a shift towards high values of the on-sky average velocity 
($\bar\mu$). Again, this shift would increase with the distance from the Sun, and this relation 
is also present in Fig.~\ref{fig7}. Despite this effect, the average velocity of the bump 
stars in NGC\,2477 ($\bar\mu\approx78$\,\kms) is exceedingly high for a $\sim10^3\,\ms$ cluster, 
to the point that these stars - if they really belong to NGC\,2477 - are not gravitationally bound 
to the cluster. This suggests that the high-velocity bump may be an artefact produced by the RL deconvolution (when applied to asymmetric profiles) or, least probably, unaccounted for field contamination.

Given the above caveat, it is not possible to disentangle a physical relation between \st\ 
and cluster age (Fig.~\ref{fig7} - right panels) from the observational limitation.

\section{Summary and conclusions}
\label{Conclu}

A crucial point in understanding a cluster's dynamical stage is the derivation of 
kinematical parameters for its member stars, usually by means of proper motions obtained
in public databases. However, uncertainties associated with ground-based proper motion 
measurements are usually large, and their effect should be properly taken into account 
when building PM profiles for determining the velocity dispersion.

In this paper we investigate the above issue using the relatively nearby and populous 
open clusters NGC\,1039 (M\,34), NGC\,2477, NGC\,2516, NGC\,2682 (M\,67), and NGC\,7762 
as test cases. Their PM components have been obtained in UCAC3. 

Rather than working with PM histograms, we build PMDFs for the cluster and field stars,
taking the $1\sigma$-PM uncertainties into account. In short, {\em (i)} we use the CMD
morphology for establishing the colour and magnitude ranges of the probable member stars,
{\em (ii)} these stars are used to build the RDP, which provides the cluster structural
parameters and allows to define the comparison field, {\em (iii)} we define a grid of PM 
bins of variable size that spans the full range of values of \mua, \mud, and \mut, {\em (iv)} 
considering the cluster and field stars separately we compute the probability that the PM 
measurements of a given star corresponds to any bin, {\em (v)} the field-star PMDF is 
subtracted from the cluster PMDF and, {\em (vi)} we take the intrinsic PM-error distribution 
function as the PSF to be used in the Richardson-Lucy deconvolution approach. 

The main result of our approach is that the deconvolved PMDFs are well represented by Gaussians 
with dispersions lower than the observed ones by a factor of 4-10. Besides the main component, 
the deconvolution revealed structure in the profiles of NGC\,2477 and NGC\,2682 in the form of 
a second - and less populous - distribution shifted towards higher average velocities, which may 
originate from cluster merger, large-scale mass segregation or, least probably, binaries. 
The secondary bump in NGC\,2477 consistently appears in PMDFs built with stars brighter and 
fainter than $J=12.5$, which suggests that it is not related to incompleteness and/or faint 
star-PM measurement. However, we cannot exclude the possibility that the bump is an 
artefact produced by the RL deconvolution when applied to strongly asymmetric profiles such
as those of NGC\,2477 and NGC\,2682. NGC\,1039 and NGC\,2516, the nearest OCs analysed 
here, end up with deconvolved dispersions compatible with those expected of bound OCs of 
$\sim10^3$\,\ms. We also detect an increase of the velocity dispersion and average velocity 
with distance from the Sun, which is probably due to a similarly limited time-base used for 
measuring the proper motions among different star clusters.

In recent years our group has given particular attention to the investigation of OCs 
by means of analytical tools that produce field-star decontaminated CMDs and RDPs. 
These tools have proved essential for a constrained analysis of OCs characterised by
a range of parameters (age, distance from the Sun, reddening, etc), and located on a 
wide variety of environments (e.g. \citealt{Teu34}; \citealt{BB07}; \citealt{N4755}).
The present paper links the classical CMD and RDP analyses with a novel approach for 
dealing with proper motions and the respective errors.

\section*{Acknowledgements}
We thank the referee, Thijs Kouwenhoven, for interesting comments and suggestions.
We acknowledge support from the Brazilian Institution CNPq.
This publication makes use of data products from the Two Micron All Sky Survey, which
is a joint project of the University of Massachusetts and the Infrared Processing and
Analysis Centre/California Institute of Technology, funded by the National Aeronautics
and Space Administration and the National Science Foundation. We also employed the WEBDA 
database, operated at the Institute for Astronomy of the University of Vienna.

\appendix
\section{Some aspects related to the RL deconvolution}
\label{Conv}

\subsection{PMDF resolution}
\label{PMDFres}

For preserving the profile resolution especially around the PMDF peak (where the number density
of stars is high) and, at the same time, keep acceptable error bars (especially towards the PMDF 
wings where the number density is low), we adopt a variable distribution of bin sizes $\delta_{pm}$, 
where $pm$ represents any component among (\mua, \mud, \mut) for building the PMDFs. 

After determining where the PMDF peak occurs, we use $\delta_{pm} = 0.25\,\mas$ for shifts 
with respect to the peak in the range $|\Delta_{pm}|=0-10\,\mas$, $\delta_{pm}=1\,\mas$ for 
$|\Delta_{pm}|=10-30\,\mas$, $\delta_{pm}=5\,\mas$ for $|\Delta_{pm}|=30-50\,\mas$, and 
$\delta_{pm} = 10\,\mas$ for $\Delta_{pm}>50\,\mas$ and $\Delta_{pm}<-50\,\mas$. Additionally, 
this procedure also has the advantage of reducing computation time.

\subsection{Deconvolution convergence}
\label{DecConv}

The rather slow convergence rate of the RL deconvolution method has been extensively
discussed in the literature (e.g. \citealt{Bi94}; \citealt{Vio05}). However, it is
easy to implement, reliable, and the quality of the outputs can be followed at each
iteration.

\begin{figure}
\resizebox{\hsize}{!}{\includegraphics{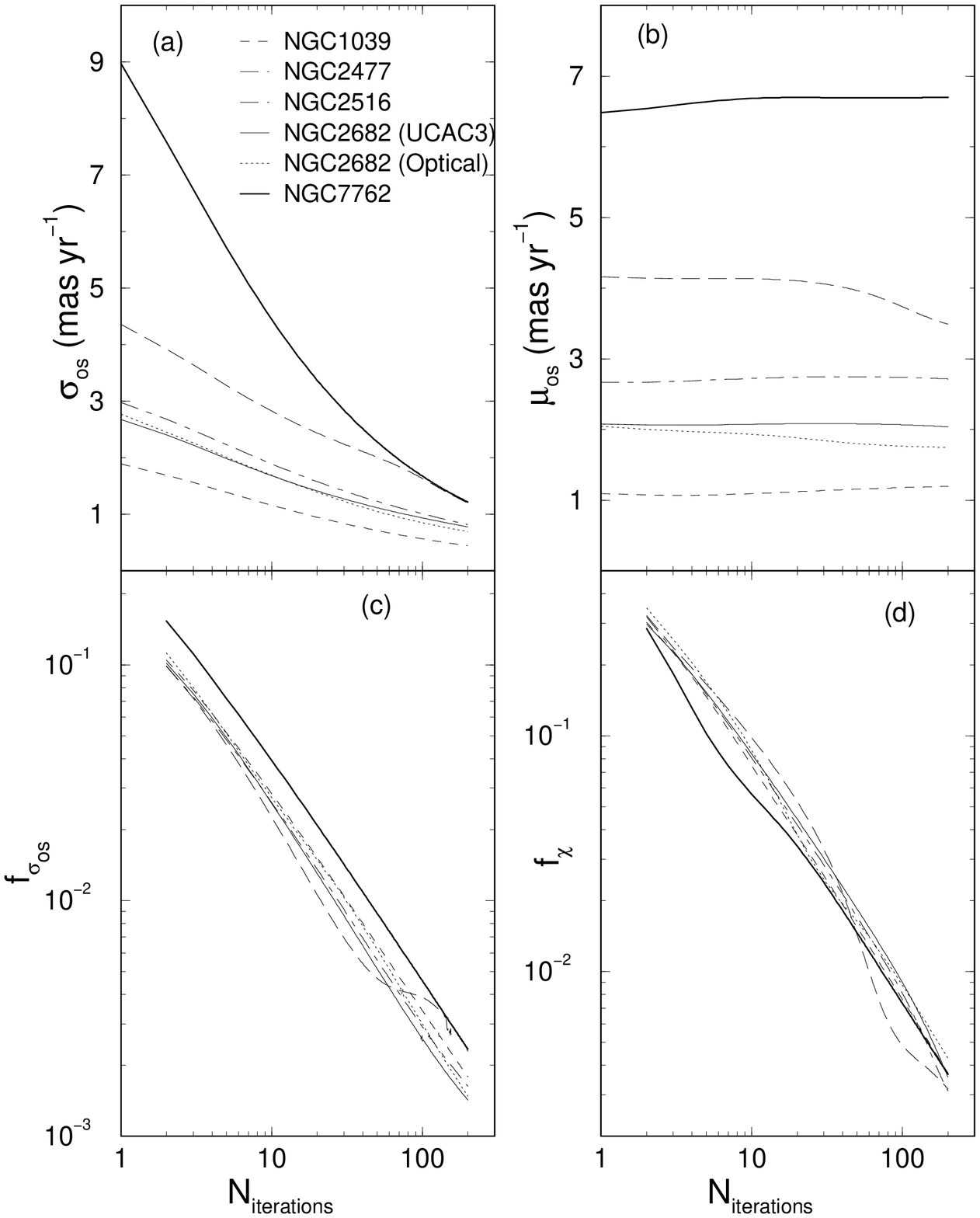}}
\caption[]{Top: variation with the number of deconvolution iterations of the on-sky velocity 
dispersion \st\ (left) and average velocity $\bar\mu$ (right). Bottom: fractional
variation of \st\ (left) and $\chi$ (right). The curves correspond to the $0\arcmin-20\arcmin$
PMDFs, except for the optical PMDF of NGC\,2682 built for $0\arcmin-15\arcmin$.}
\label{fig8}
\end{figure}

We follow in Fig.~\ref{fig8} the changes on \st\ and $\bar\mu$ with the number of deconvolution
iterations ($N_{it}$). While the average velocity (panel b) changes little over the 200 
iterations applied here, \st, on the other hand, decreases systematically with $N_{it}$ (panel a) but 
with a rate that begins to flatten for $N_{it}\ga100$. Clearly, \st\ could decrease somewhat more 
for $N_{it}>200$. However, as implied by the rate of change $f_{\st}=\Delta\st/\st$ that, for 
$N_{it}=200$ has decreased to $f_{\st}\la0.003$ (panel c), it would take several hundred more iterations 
to produce a significant change in \st. Finally, the chi-square (Sect.~\ref{RLD}) fractional variation
$f_{\chi} = \Delta\chi/\chi$ also presents a steady decrease with $N_{it}$ (panel d), dropping to
$f_{\chi}\sim0.003$ (or a $\sim0.3\%$ variation) for $N_{it}=200$.

\section{Observed and field PMDFs}
\label{OFP}

\begin{figure}
\resizebox{\hsize}{!}{\includegraphics{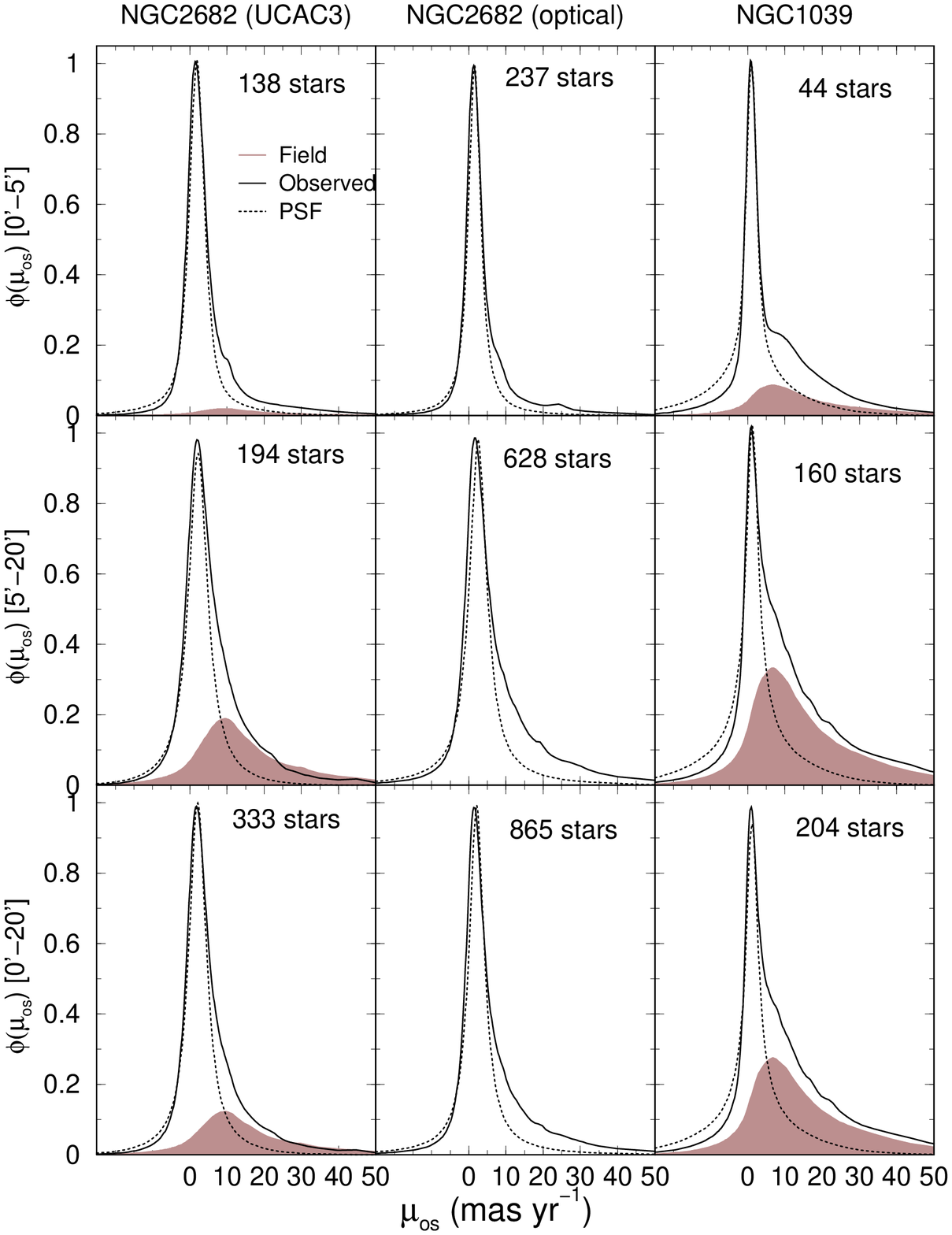}}
\caption[]{The panels show the observed, field, and intrinsic error (PSF) on-sky PMDFs
built for different cluster regions: central (top), outer (middle) and overall (bottom).
The number of member stars in each region is indicated. }
\label{fig9}
\end{figure}

\begin{figure}
\resizebox{\hsize}{!}{\includegraphics{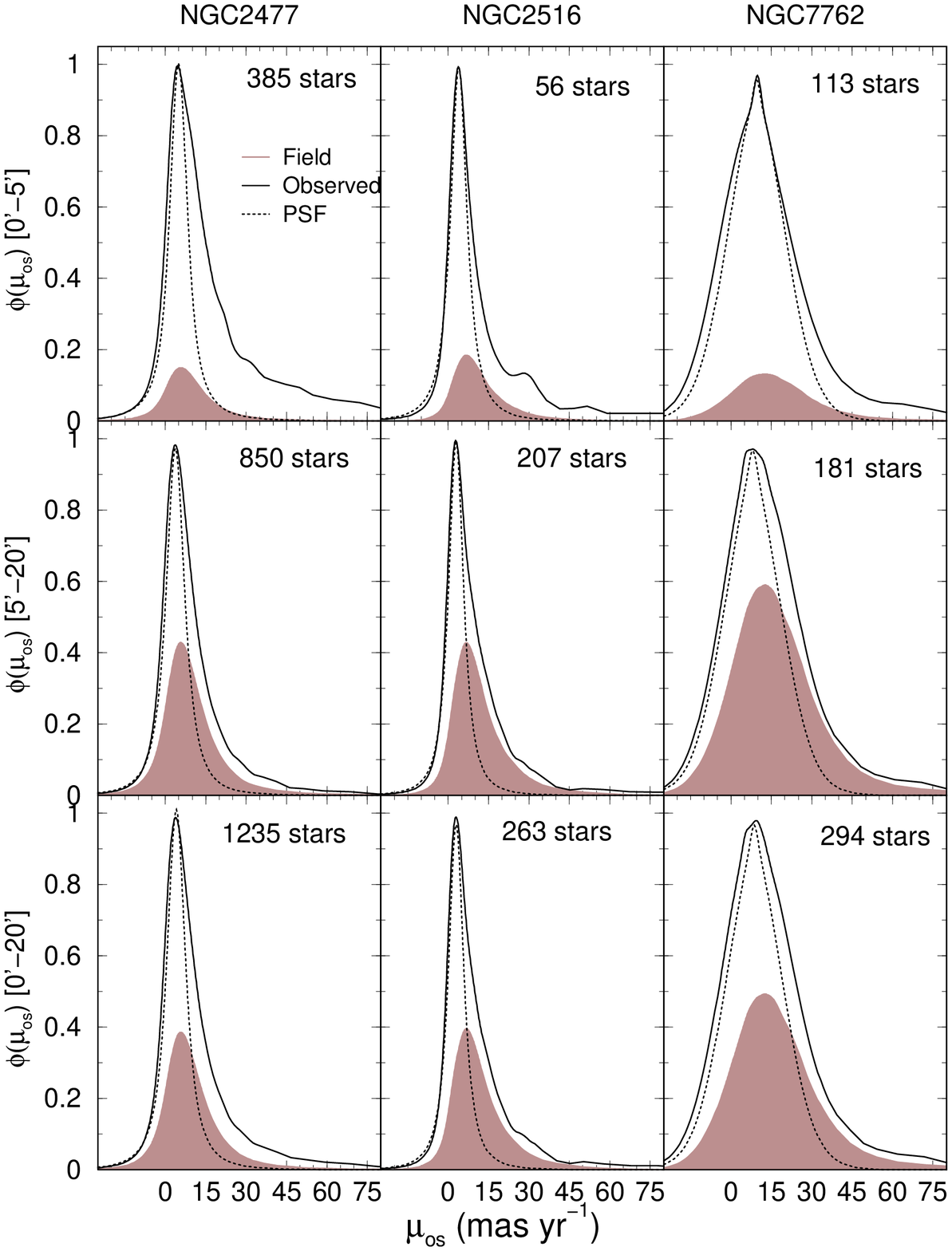}}
\caption[]{Same as Fig.~\ref{fig9} for the remaining OCs. }
\label{fig10}
\end{figure}

We show in Figs.~\ref{fig9} and \ref{fig10} the observed on-sky PMDFs for all clusters of 
our sample, built for the central, outer, and overall cluster regions. Also shown are the PMDFs
corresponding to the intrinsic error distribution (PSF) and the field. It is interesting to
note the relative contribution of the field stars PM among the different regions of the same
cluster and among the full set of clusters. This shows that the field contribution must be taken 
into account in the analysis. Also, and perhaps more importantly is the PSF width that, in some
cases, is almost as broad as the cluster PMDF.

\subsection{PMDFs in different magnitude ranges}
\label{dMagBump}

\begin{figure}
\resizebox{\hsize}{!}{\includegraphics{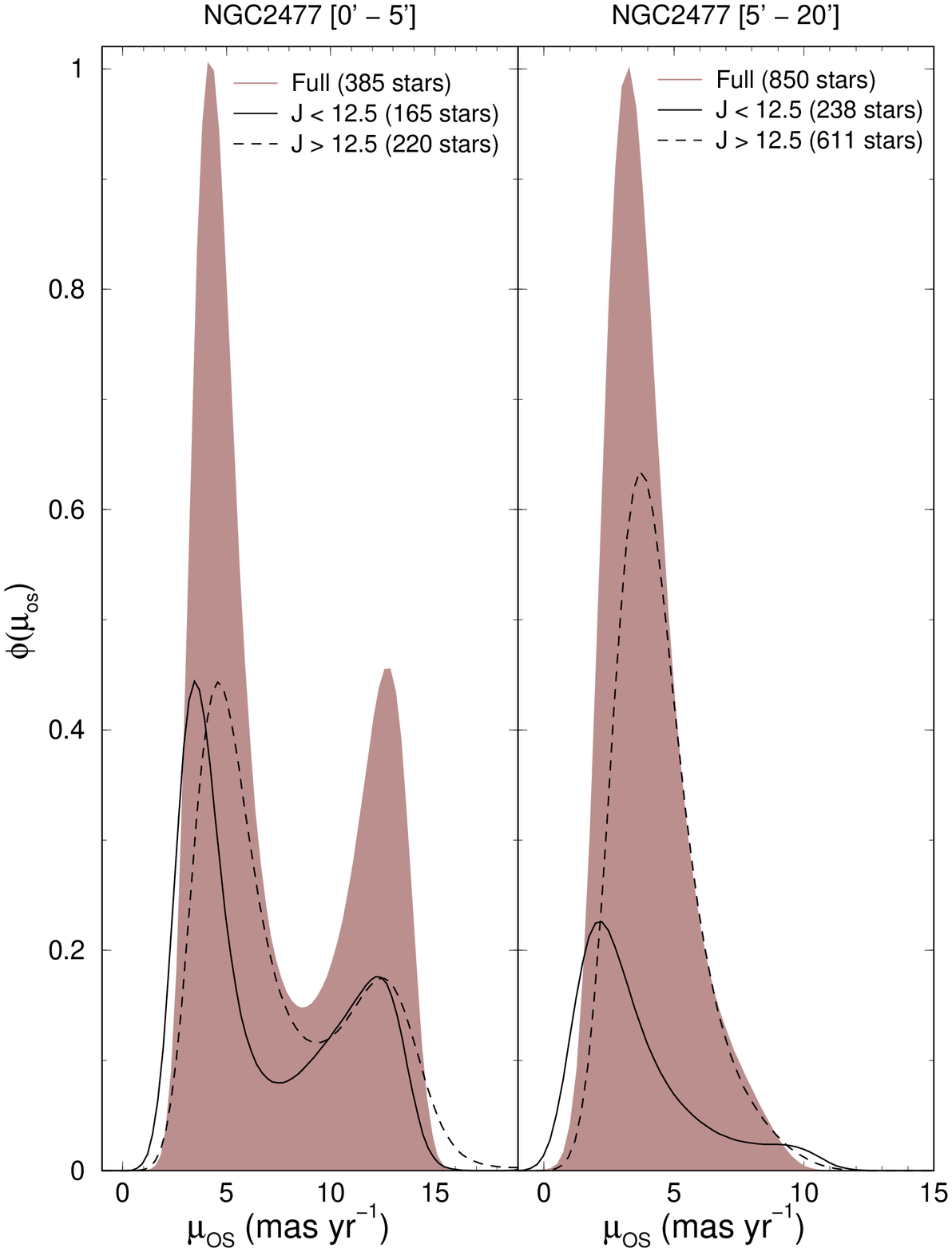}}
\caption[]{Deconvolved PMDFs for the central and outer regions of NGC\,2477, separated by
magnitude ranges. The bright (solid line) and faint (dashed) PMDFs are compared to the 
``full-magnitude'' (shaded) PMDF. The number of stars composing the PMDFs is indicated.}
\label{fig11}
\end{figure}

In Fig.~\ref{fig11} we examine the incompleteness/faint stars issue in NGC\,2477, which should be 
important especially for the central part of the cluster. Starting with the CMD of NGC\,2477 
(Fig.~\ref{fig2}), we build PMDFs separately for stars brighter and fainter than $J=12.5$. This 
magnitude boundary is adequate for characterising different types of stars and, at the same time, 
keeping a reasonable number of stars in each magnitude range. The bright and faint PMDFs are
compared to the ``full-magnitude'' range PMDF, both for the central (left panel) and outer
(right) cluster region.

\label{lastpage}
\end{document}